\documentclass[aps,pra,reprint,amsmath,amssymb,superscriptaddress]{revtex4-2}

\usepackage[utf8]{inputenc}
\usepackage[T1]{fontenc}
\usepackage{graphicx}
\usepackage[colorlinks=true,urlcolor=blue,citecolor=blue,linkcolor=blue,bookmarks=false,pdfstartview={FitH}]{hyperref}

\usepackage{color,soul}

\begin{document}
	
	\title{Anomalous phonon Gr\"uneisen parameters in semiconductor Ta$_2$NiS$_5$}
	
	\author{Mai Ye}
	\email{mai.ye@kit.edu}
	\affiliation{Institute for Quantum Materials and Technologies, Karlsruhe Institute of Technology, Kaiserstr. 12, 76131 Karlsruhe, Germany}
	\author{Tom Lacmann}
	\affiliation{Institute for Quantum Materials and Technologies, Karlsruhe Institute of Technology, Kaiserstr. 12, 76131 Karlsruhe, Germany}
	\author{Mehdi Frachet}
	\altaffiliation[Present address: ]{CEA-Leti, Université Grenoble Alpes, Grenoble, France}
	\affiliation{Institute for Quantum Materials and Technologies, Karlsruhe Institute of Technology, Kaiserstr. 12, 76131 Karlsruhe, Germany}
	\author{Igor Vinograd}
	\altaffiliation[Present address: ]{4th Physical Institute – Solids and Nanostructures, University of Göttingen, 37077 Göttingen, Germany}
	\affiliation{Institute for Quantum Materials and Technologies, Karlsruhe Institute of Technology, Kaiserstr. 12, 76131 Karlsruhe, Germany}
	\author{Gaston Garbarino}
	\affiliation{European Synchrotron Radiation Facility, BP 220, F-38043 Grenoble Cedex, France}
	\author{Nour Maraytta}
	\affiliation{Institute for Quantum Materials and Technologies, Karlsruhe Institute of Technology, Kaiserstr. 12, 76131 Karlsruhe, Germany}
	\author{Michael Merz}
	\affiliation{Institute for Quantum Materials and Technologies, Karlsruhe Institute of Technology, Kaiserstr. 12, 76131 Karlsruhe, Germany}
	\affiliation{Karlsruhe Nano Micro Facility (KNMFi), Karlsruhe Institute of Technology, Kaiserstr. 12, 76131 Karlsruhe, Germany}
	\author{Rolf Heid}
	\affiliation{Institute for Quantum Materials and Technologies, Karlsruhe Institute of Technology, Kaiserstr. 12, 76131 Karlsruhe, Germany}
	\author{Amir-Abbas Haghighirad}
	\affiliation{Institute for Quantum Materials and Technologies, Karlsruhe Institute of Technology, Kaiserstr. 12, 76131 Karlsruhe, Germany}
	\author{Matthieu Le Tacon}
	\email{matthieu.letacon@kit.edu}
	\affiliation{Institute for Quantum Materials and Technologies, Karlsruhe Institute of Technology, Kaiserstr. 12, 76131 Karlsruhe, Germany}
	
	\date{\today}
	
	\begin{abstract}
		Strain tuning is a powerful experimental method in probing correlated electron systems. Here we study the strain response of the lattice dynamics and electronic structure in semiconductor Ta$_2$NiS$_5$ by polarization-resolved Raman spectroscopy. We observe an increase of the size of the direct semiconducting band gap. Although the majority of the optical phonons show only marginal dependence to applied strain, the frequency of the two B$_{2g}$ phonon modes, which have quadrupolar symmetry and already anomalously soften on cooling under zero strain, increases significantly with tensile strain along the $a$ axis. The corresponding Gr\"uneisen parameters are unusually large in magnitude and negative in sign. These effects are well captured by first-principles density functional theory calculations and indicate close proximity of Ta$_2$NiS$_5$ to a structural instability, similar to that encountered in excitonic insulator candidate  Ta$_2$NiSe$_5$.
	\end{abstract}
	
	\maketitle
	
	\section{Introduction\label{sec:Intro}}
	
	A macroscopic number of bound pairs of fermions, formed at low temperatures due to attractive interaction, result in quantum phases with superconductivity~\cite{BCS} or superfluidity~\cite{Anderson}. Excitonic insulators (EIs) belong to such quantum phases, in that the Coulomb attraction in a semimetal or in a narrow-gap semiconductor leads to an instability towards the formation of electron-hole pairs (excitons)~\cite{Ex1967,jerome1967,Ex1968} and these bosonic quasiparticles can collectively establish a coherent state and open an interaction-induced gap~\cite{Ex1970}. Furthermore, if the participating electron and hole states belong to bands of different symmetry, the resulting EI breaks the crystal lattice symmetry.
	
	The material family Ta$_2$Ni(Se$_{1-x}$S$_x$)$_5$ serves as a suitable platform to investigate the rich physics of EIs. On one end of its phase diagram, Ta$_2$NiSe$_5$ is largely believed to be a semimetal at high temperature~\cite{ARPES2020,Pavel2021a} which has been proposed to host an EI state~\cite{ARPES2009} below an orthorhombic-to-monoclinic structural phase transition around 326\,K~\cite{Structure1986,Transport2017}. 
	This transition reduces the point-group symmetry from $D_{2h}$ to $C_{2h}$ by breaking two mirror symmetries, and by group theory the corresponding order parameter is of B$_{2g}$ symmetry. It is accompanied with the opening of a gap in the electronic dispersion and a band flattening~\cite{ARPES2009}, but there is generally a controversy on whether these features can be effectively attributed to the condensation of excitons~\cite{ARPES2013,ARPES2014,Mazza_PRL2020,ARPES2021,Fast2023a}, or they are only a consequence of the lattice symmetry lowering and electron-phonon coupling~\cite{ARPES2020,Fast2023b,ARPES2023}.
	
	The transition temperature is suppressed by S substitution~\cite{Transport2017,Pavel2021,Mai2021}, and at the other end of the phase diagram, the iso-structural compound Ta$_2$NiS$_5$ is a semiconductor with a band gap exceeding the exciton binding energy, henceforth forbidding the EI instability~\cite{Structure1986,ARPES2019S,Chen2023}.
	In contrast with Ta$_2$NiSe$_5$, the properties of semiconducting Ta$_2$NiS$_5$ have been relatively little explored.
	It is for instance unclear whether the aforementioned structural transition is completely suppressed. This has been suggested in recent x-ray diffraction experiments~\cite{Chen2023}, but Raman scattering investigation~\cite{Mai2021} has interpreted a temperature-dependent breaking of the phonon polarization selection rules below 120 K as evidence for such transition. This view is further supported by first-principles calculations~\cite{DFT2021} indicating that in both Ta$_2$NiSe$_5$ and Ta$_2$NiS$_5$, the lattice symmetry is broken by phonon instabilities, emphasizing the importance of lattice degrees of freedom in shaping the phase diagram of this family of materials.
	
	In this work, we tackle this issue from a different perspective, probing the response of the phononic and electronic degrees of freedom of Ta$_2$NiS$_5$ to uniaxial strain. Strain tuning has proven particularly insightful in studying the electronic properties of unconventional low-~\cite{Hicks2014, Steppke2017} or high-temperature superconductors~\cite{MLT2018,MLT2021,vinograd2023}, ferromagnets~\cite{Najev2022}, antiferromagnets~\cite{Keimer2022}, and topological systems~\cite{Joshua2019}. Here we combine it with Raman spectroscopy and first-principles calculations to reveal the highly anomalous behavior of two phonon modes of B$_{2g}$ symmetry involving shear vibrations of the Ta atoms along the crystallographic a-direction. We show that this behavior arises independently from changes of the band gap size and points towards an intrinsic structural instability of the orthorhombic unit cell upon volume reduction. 
	More generally, our results demonstrate the possibility of manipulating the electronic band gap in the Ta$_2$Ni(Se$_{1-x}$S$_x$)$_5$ family using strain, which offers interesting perspectives to approach the still elusive EI regime.
	
	\section{Method\label{sec:Exp}}
	High-quality single crystals of Ta$_2$NiS$_5$ were grown using chemical vapor transport~\cite{SOM}. The composition of the obtained crystals was performed using energy dispersive x-ray spectroscopy (EDX) in a COXEM EM-30plus electron microscope equipped with an Oxford Silicon-Drift-Detector (SDD) and AZtecLiveLite-software package.
	
	Strain-dependent Raman-scattering experiments were performed with a Horiba Jobin-Yvon LabRAM HR evolution spectrometer.
	We used a He-Ne laser (632.8\,nm) with less than 1\,mW power that was focused on the sample with a x50 magnification objective. The laser spot size was around 5\,$\mu$m in diameter. One notch filter and two Bragg filters were used in the collection optical path to filter the elastically scattered light.
	Spectra were recorded with 1800\,mm$^{-1}$ (0.6\,cm$^{-1}$ spectral resolution) and 600\,mm$^{-1}$ (1.6\,cm$^{-1}$ spectral resolution) gratings and a liquid-nitrogen-cooled CCD detector. All spectra of Raman response were corrected for the instrumental spectral response and Bose factor.
	Two polarization configurations were employed to probe excitations in different symmetry channels, namely the $aa$ and $ac$ configurations in which the incident light polarization lies along the crystallographic $a$ direction, and scattered one along $a$ or $c$ direction, respectively. According to Raman selection rules~\cite{Hayes2004}, in the orthorhombic phase (point group D$_{2h}$), the $aa$ and $ac$ geometries probe the A$_{g}$ and B$_{2g}$ Raman signal, respectively.
	Samples with a cleaved crystallographic $ac$ plane were glued to Razorbill CS220T strain cell by Loctite Stycast 2850FT epoxy with CAT 24LV as the epoxy catalyst. The sample surfaces were examined under polarized light to find an area without intrinsic strain. We verified that the process of mounting the sample onto the strain cell does not introduce additional stress by comparing the Raman spectra of unmounted and mounted samples. The strain transmission was estimated from XRD measurement as detailed in the supplemental material~\cite{SOM}; the corresponding data measured at European Synchrotron Radiation Facility (ESRF) ID15B beamline are available from the ESRF Data Portal~\cite{ESRF}.
	
	Density functional theory (DFT) calculations were performed using the mixed-basis pseudopotential method \cite{meyer97} to assess the effect of uniaxial strain on the electronic structure and Raman active modes~\cite{SOM}.
	The exchange-correlation functional was represented by the generalized gradient approximation (GGA) \cite{perde96}. The DFT\,+\,U approach was applied with U=10\,eV for the Ni $d$ states. We performed all calculations in the room-temperature orthorhombic structure (space group $Cmcm$).
	Lattice constants were taken from Sunshine et al. \cite{sunsh85}, $a = 3.415$~\AA, $b = 12.146$~\AA, $c = 15.097$~\AA. The relatively small tensile strain was modeled by scaling $a$ only, ignoring any elastic reaction of the $b$ or $c$ axes.
	For each structure, atomic positions were relaxed until forces were smaller than 2.6~meV/\AA.
	Raman-mode frequencies were calculated using the linear response, or density-functional perturbation theory (DFPT), implemented in the mixed-basis pseudopotential method \cite{heid99}.
	
	\section{Results and Discussion\label{sec:Res}}
	
	\subsection{Strain-free phonon spectra}
\begin{figure}
		\includegraphics[width=0.98\linewidth]{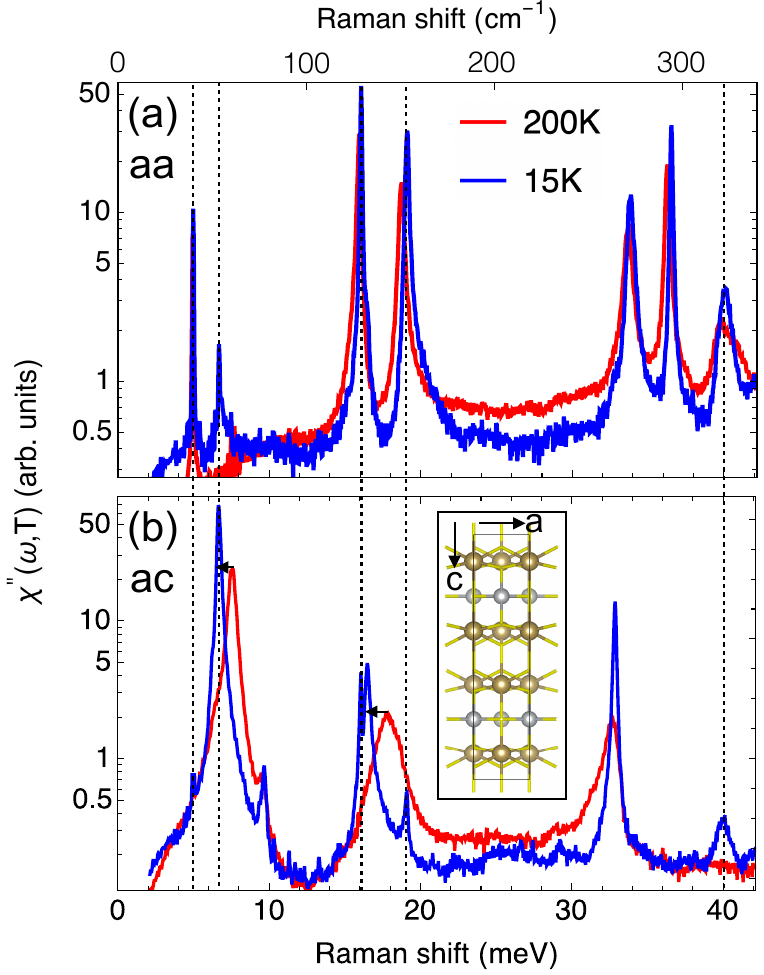}
		\caption{\label{fig:OP} Log-scale Raman response $\chi^{\prime\prime}$ in the $aa$ (a) and $ac$ (b) scattering geometry, respectively. The corresponding light polarizations are illustrated at top left of each panel. The spectral features of phonon-mode "leakage" are labelled by dashed vertical lines. The anomalous softening of the two lowest-energy B$_{2g}$ phonons is labelled by solid horizontal arrows in panel (b). The inset in panel (b) is a sketch of the crystal structure with the a- and c-axes labelled. The tantalum atoms are shown in brown and the nickel atoms are shown in grey. The sulfur atoms are omitted for clarity.}
	\end{figure}

	In agreement with previous Raman-scattering investigation~\cite{Mai2021}, we observe a "leakage" of certain phonon modes at low temperatures. This is illustrated by the Raman spectra measured at 200\,K and 15\,K in two configurations, as shown in Fig.~\ref{fig:OP}. Up to 40\,meV, and in agreement with the selection rules of the orthorhombic phase, six A$_{g}$ Raman phonons are observed in the $aa$ scattering geometry and three B$_{2g}$ modes in the $ac$ scattering geometry. At low temperature, on the contrary, the lowest-energy B$_{2g}$ mode becomes apparent ("leaks") in the $aa$ scattering geometry, breaking down the aforementioned selection rules. Similarly, four A$_{g}$ modes "leak" in the spectra recorded in the $ac$ scattering geometry at low temperature. Such leakage strongly suggests a reduction of the D$_{2h}$ point group symmetry of the orthorhombic phase and, akin to Ta$_2$NiSe$_5$~\cite{Structure1986}, the most natural point group candidate would correspond to a monoclinic C$_{2h}$ phase that breaks the $m_a (a \rightarrow -a)$ and $m_c (c \rightarrow -c)$ mirror symmetries.
	
	The phonon leakage is an indication of symmetry breaking, and the phonon softening suggests a tendency towards an orthorhombic-to-monoclinic structural distortion. A recent XRD experiment finds no clear evidence for a structural transition~\cite{Chen2023}. Our in-house XRD investigation could not reveal any signature of such transition down to 80 K~\cite{SOM} and our refinement of the crystal structure of Ta$_2$NiS$_5$ at 15\,K with data taken at the synchrotron is not yet fully conclusive~\cite{SOM}. As a result additional studies are called for to clarify this issue.
	
	Nonetheless, another noteworthy feature of the Raman spectra of Ta$_2$NiS$_5$ is the anomalous softening of the two lowest-energy B$_{2g}$ modes (hereafter labelled B$_{2g}^{(1)}$ and B$_{2g}^{(2)}$) on cooling. As previously noted~\cite{Mai2021}, this behavior is inconsistent with the conventional anharmonic decay of phonons in solids and might indicate the proximity of a phase transition.
	Because of the lattice contraction upon cooling, the corresponding Gr\"uneisen parameter $\gamma_i = -(\Delta \omega_i/\omega_i)/(\Delta V/V)$ of these phonons is negative, which indicates that the orthorhombic structure is generally unstable against a reduction of the unit cell volume.
	Interestingly, the order parameter of the orthorhombic-to-monoclinic transition in Ta$_2$NiSe$_5$ has the B$_{2g}$ symmetry and the vibration pattern of the two lowest-energy B$_{2g}$ modes, which couple to the overdamped excitonic mode~\cite{Pavel2021}, consists of a shear displacement of the Ta along the a-direction, reminiscent of the structural distortion (in contrast with the B$_{2g}^{(3)}$ mode which involves mostly S vibrations and behaves normally). Nevertheless, the relevant B$_{2g}$ modes in Ta$_2$NiSe$_5$ exhibit no softening behavior~\cite{Pavel2021,Mai2021}, consistent with the view that the transition is driven by an excitonic instability.
	Taken together, these observations indicate that lattice strain tuning might be a promising tuning knob of the electronic properties of this system, as discussed below.
	
	\subsection{Strain-dependent Raman spectra}
	Next, we investigate the strain response of the system. Because of the anisotropic structure of Ta$_2$NiS$_5$, and the fact that the putative structural distortion of the material is of B$_{2g}$ symmetry which transforms as an $ac$ quadrupole, we choose to focus on the spectral response of tensile strain along the crystallographic $a$ direction, which is also favoured by the morphology of the single crystals.
	
\begin{figure}
		\includegraphics[width=0.98\linewidth]{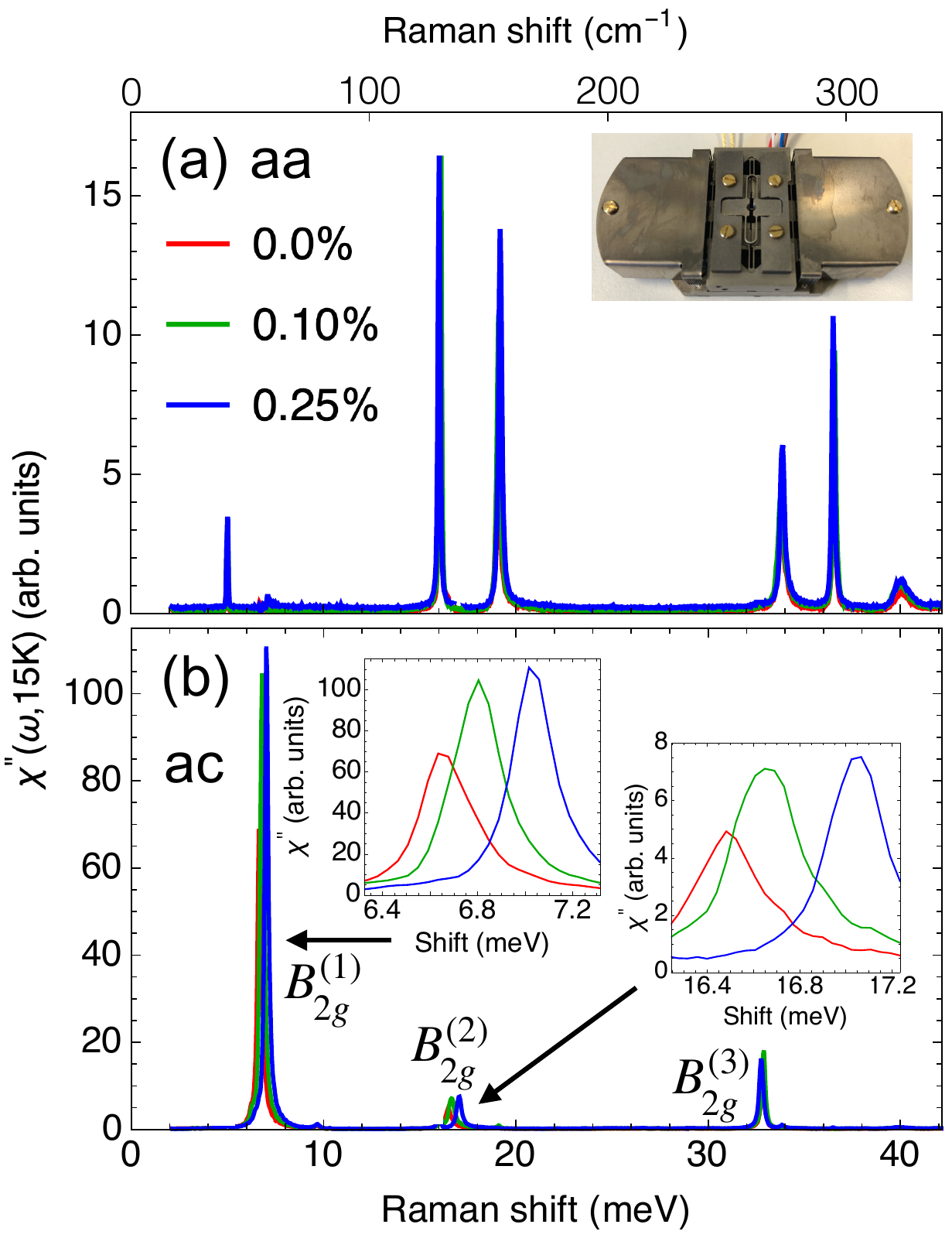}
		\caption{\label{fig:Pola} Raman response $\chi^{\prime\prime}$ measured under various $a$ axis tensile strain at 15\,K in the (a) aa and (b) ac scattering geometries. The inset in (a) exhibits the Razorbill strain cell CS220T used for the Raman experiments. The insets in (b) show the strain dependence of the two lowest-frequency B$_{2g}$-symmetry phonon modes.}
	\end{figure}

	In Fig.~\ref{fig:Pola}, we show the phonon spectra measured at 15\,K as function of $a$-axis tensile strain up to 0.25\%. As previously explained, in the monoclinic phase, all the modes observed in $aa$ [Fig.~\ref{fig:Pola} (a)] or $ac$ [Fig.~\ref{fig:Pola} (b)] configurations have the A$_g$ symmetry of C$_{2h}$ group. However, we observe a clear qualitative difference of the strain response for the modes which have larger intensity in the $aa$ configuration (which had the A$_g$ symmetry of D$_{2h}$ group in the high-temperature orthorhombic phase) and those seen in the $ac$ scattering geometry that had B$_{2g}$ character at high temperature. For clarity, we continue to label these phonons according to their high temperature symmetry. Strikingly, the B$_{2g}^{(1)}$ and B$_{2g}^{(2)}$ harden significantly and their lineshape narrows with increasing tensile strain [Insets of Fig.~\ref{fig:Pola} (b)]. This behavior starkly contrast with the behavior of the B$_{2g}^{(3)}$ mode or with the observation in the $aa$ configuration, where we can barely detect a change in the frequency or linewidth of the A$_{g}$ phonons in the investigated strain range.
	
\begin{figure}
		\includegraphics[width=0.98\linewidth]{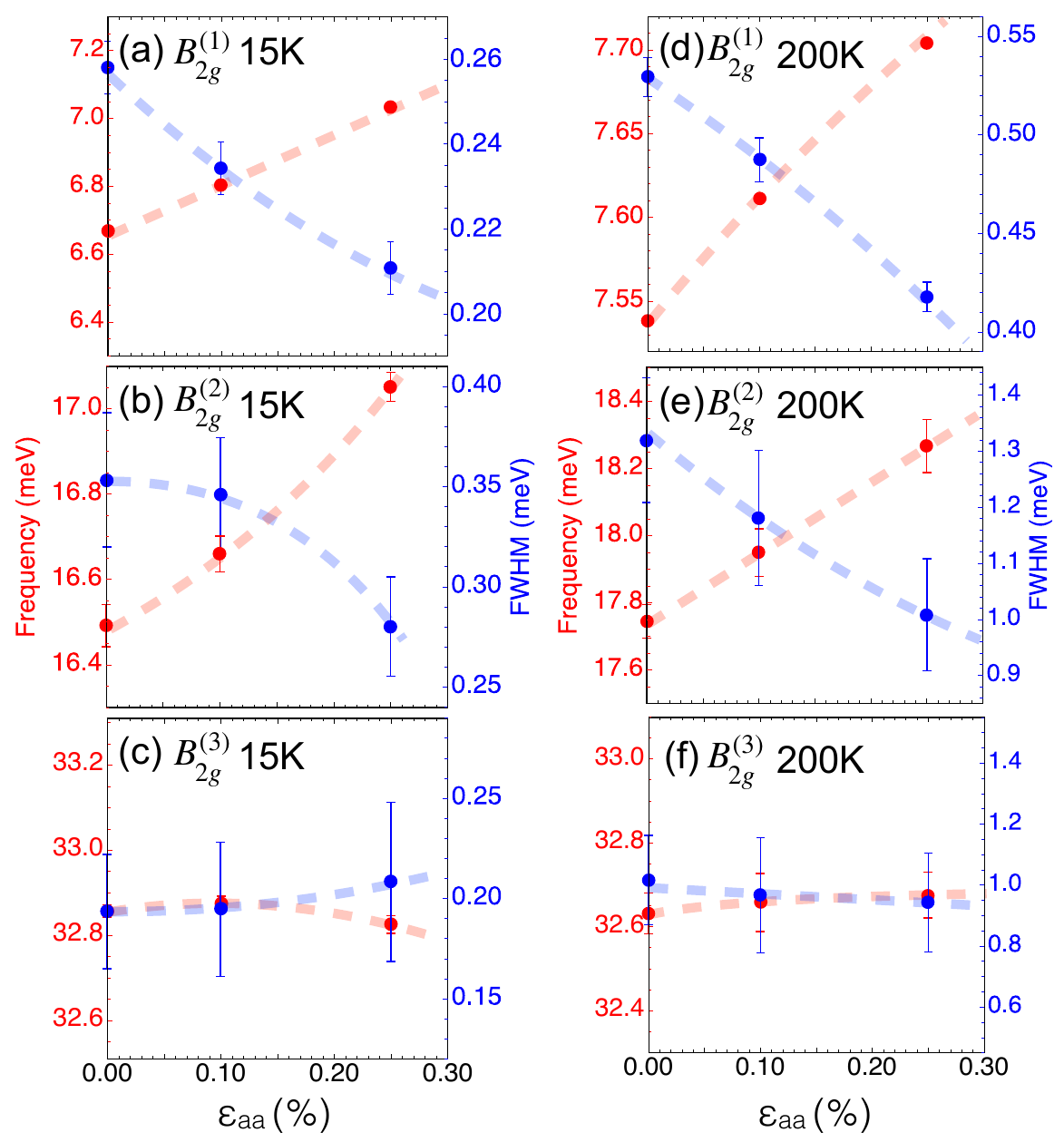}
		\caption{\label{fig:Para}Tensile strain ($\epsilon_{aa}$) dependence of the frequency and full width at half maximum (FWHM) for the three B$_{2g}$ phonon modes at 15\,K (a-c) and 200\,K (d-f). The dashed lines are a guide to the eyes.}
	\end{figure}

	In order to quantify the magnitude of these effects, we have fitted the different modes using standard Lorentzian lineshape. The strain dependence of the frequency and linewidth at 15\,K for the three B$_{2g}$ phonon modes are shown in Fig.~\ref{fig:Para} (a-c). With 0.25\,\% strain, the relative frequency shift for the B$_{2g}^{(1)}$, B$_{2g}^{(2)}$, and B$_{2g}^{(3)}$ modes are 5.5, 3.4, and 0.1\,\%, respectively. The relative frequency changes for the A$_{g}$ modes are all less than 0.5\,\%. Besides the significant hardening of frequency, the B$_{2g}^{(1)}$ and B$_{2g}^{(2)}$ modes exhibit a 20\% reduction of linewidth. This effect is completely absent for the other modes, whose linewidth is nearly strain-independent. The behaviors of the linewidth demonstrate that within the probed volume the strain is homogeneous, because inhomogeneity would result in lineshape broadening~\cite{RamanStrainW}.
	
	The relative frequency change, 5.5\,\% and 3.4\,\%, of the two lowest-energy B$_{2g}$ phonon modes at 0.25\% strain is anomalously large compared to other systems. To fix ideas regarding the amplitude of this effect, we provide hereafter a few examples from the recent litterature for comparison. We first consider the silicon T$_{2g}$-symmetry phonon mode which has a frequency of around 520\,cm$^{-1}$. Under as large as 8.5\,\% strain induced in silicon membranes by ion implantation, the magnitude of the relative frequency change is 4.9\,\%~\cite{Silicon2021}, which is in between the values 5.5\,\% and 3.4\,\% we found. We then consider some two-dimensional systems whose strain response is actively studied~\cite{Review2020}. The phonon modes of monolayer transition-metal-dichalcogenides WSe$_2$ and WS$_2$ only show around 2\% frequency shift at 2.85\% tensile strain~\cite{RamanStrainW}, and CrSBr flake exhibits about 2\% frequency change at 1.7\% tensile strain~\cite{RamanStrainC}. These ratios between the frequency change and strain are closer to the strain effect observed for the $A_g$ phonon modes of Ta$_2$NiS$_5$, and generally expected from Gr\"uneisen's law, with Gr\"uneisen parameters of individual phonons lying in the range of a few units for most systems semiconductor and metallic systems~\cite{Ashcroft1976}. Here and in agreement with the temperature dependent experiment discussed in the previous section, the Gr\"uneisen parameters for the B$_{2g}^{(1)}$ and B$_{2g}^{(2)}$ phonon modes in Ta$_2$NiS$_5$ are not only much larger in magnitude but also anomalously negative, reaching -22 and -13.6, respectively.
	
\begin{figure}
		\includegraphics[width=0.9\linewidth]{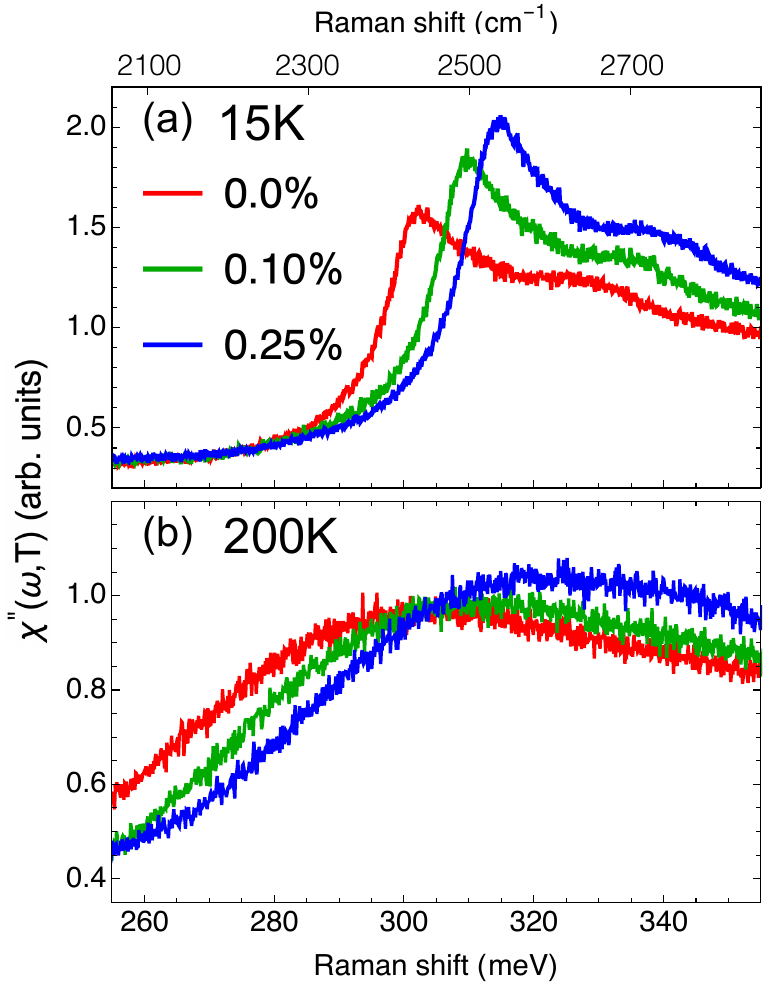}
		\caption{\label{fig:Gap}High-energy Raman response $\chi^{\prime\prime}$ measured under various $a$ axis tensile strain at (a) 15\,K and (b) 200\,K in the $ac$ scattering geometry.}
	\end{figure}

	To find out whether the anomalous behavior of the B$_{2g}^{(1)}$ and B$_{2g}^{(2)}$ modes could be related to strain-induced changes in the electronic structure, we also studied the strain-dependence of the electronic excitations. As previously reported~\cite{Pavel2021}, the presence of a direct semiconducting gap suppresses every low-energy electronic excitation in Ta$_2$NiS$_5$. The lowest interband transition manifests itself already at 200\,K in the form of a broad excitation centered around 300\,meV [Fig.~\ref{fig:Gap} (b)]. At 15\,K, the spectrum sharpens significantly and the overall profile changes [Fig.~\ref{fig:Gap} (a)]. We do not see the B$_{2g}$-symmetry exciton mode as clearly sharp and strong as in Ref.~\cite{Pavel2021}, presumably because of the narrow resonance profile of this feature (our measurement is performed with 633\,nm laser excitation rather than with 647\,nm line used in Ref.~\cite{Pavel2021}). The spectra are however perfectly consistent with previous reports, showing as well a weaker spectral feature at around 330\,meV, which has previously been interpreted as the second state of the Rydberg series~\cite{Pavel2021}. Under applied $a$-axis stress, a continuous hardening of the electronic excitation spectra arises both at high and low temperatures. With 0.25\% strain, a relative blue shift of the excitation spectra amounting to 4.3\% (from 302 to 315\,meV) is observed. This 4.3\% change of band gap at 0.25\% strain is now well comparable to the behaviors encountered in other semiconducting systems such as bilayer MoS$_2$ under tensile strain~\cite{RamanStrainM1}, or trilayer MoS$_2$ under various biaxial compressive strains~\cite{RamanStrainM2}. Qualitatively, we can associate this behavior with a reduction, under tensile strain along the crystallographic $a$ direction, of the overlap between Ta 5d orbitals (so their hybridization), on the one hand, and between the Ni 3d orbitals on the other hand, which yields an overall reduction of the bandwidth of the respective bands across the direct semiconducting gap, naturally yielding its increase. At the same time, the concomitant anomalous hardening of the two B$_{2g}^{(1)}$ and B$_{2g}^{(2)}$ phonon frequencies indicates an increased stability of the structure at larger unit cell volumes, hinting towards an intrinsic structural instability of the shrunk unit cell (\textit{e.g.} under the effect of cooling).
	
	\subsection{First-principles calculations and discussion}
	
	To gain further insights, it is instructive to compare our experimental results with those from first principle calculations. In Table~\ref{tab:DFT1}, we compare the experimentally measured energy of three optical phonon modes and the band gap with the calculated values (note that experimental lattice parameters are used for the calculation~\cite{SOM}). Interestingly, within our computational approach, the eigenvalue of the dynamical matrix for the B$_{2g}^{(1)}$ mode is imaginary, and the structure therefore unstable against the eigendisplacement of this mode, which precisely involve the shear motion of the Ta ions along the a axis that would yield a monoclinic distortion akin to the Se compound.
	Leaving this aside, we can look at the electronic structure. As discussed in the supplemental material~\cite{SOM} and in agreement with previous computational work~\cite{DFT2021}, correlation corrections for the Ni $d$ states need to be introduced in the framework of the GGA\,+\,$U$ approach to open a direct semiconducting gap in the electronic structure. A 126\,meV gap is obtained for $U$\,=\,10\,eV, which is significantly lower than the experimental value indicating that non-local self energy effects, especially that of Ta and Ni orbitals near the Fermi level, should be taken into account to reproduce the realistic band gap~\cite{DFT2021}. Nevertheless, the relative magnitude of the strain-induced effects on the gap is correctly captured by our calculation, validating the qualitative picture drawn earlier.
	
	We also find that the calculated energy of the phonon modes is in good agreement with the experimental values, and the absolute renormalization of the B$_{2g}^{(2)}$ frequency with strain as well as the absence of renormalization for the A$_{1g}^{(1)}$ and B$_{2g}^{(3)}$ (and so essentially their respective Gr\"uneisen parameters) are perfectly captured by the calculation. As already emphasized, the B$_{2g}^{(2)}$ (and B$_{2g}^{(1)}$) mode involves a shear motion of Ta ions along a, a pattern that was previously found theoretically to strongly modulate the semiconducting gap size~\cite{DFT2021}. This raises the question to know whether the anomalous behavior of this mode should then be attributed to the strain-induced change of the semiconducting gap, or whether it rather arises only from lattice stretching.
	
	\begin{table}
		\caption{\label{tab:DFT1}Comparison between the experimentally measured (Exp.) energy values and the calculated (Cal.) ones for selected phonon modes and band gap at 15\,K. The A$_{1g}^{(1)}$ mode refers to the A$_{1g}$ phonon with the lowest energy. The experimentally measured band gap is defined as the energy at which the interband excitations have the maximal intensity; the calculated value corresponds to a direct gap at $\Gamma$ point. The values in the brackets give the change compared to the corresponding zero-strain value. The unit for energies is meV.}
		\begin{ruledtabular}
			\begin{tabular}{ccccccc}
				Strain&\multicolumn{2}{c}{0\,\%}&\multicolumn{2}{c}{0.1\,\%}&\multicolumn{2}{c}{0.25\,\%}\\
				&Exp.&Cal.&Exp.&Cal.&Exp.&Cal.\\
				\hline
				Gap&302&126&309&132&315&146\\
				&&&(+7)&(+7)&(+13)&(+20)\\
				\hline
				A$_{1g}^{(1)}$&4.99&4.79&5.01&4.78&5.01&4.81\\
				&&&(+.02)&(-.01)&(+.00)&(+.02)\\
				B$_{2g}^{(2)}$&16.49&15.18&16.66&15.41&17.05&15.73\\
				&&&(+.17)&(+.23)&(+.56)&(+.55)\\
				B$_{2g}^{(3)}$&32.86&32.46&32.87&32.50&32.83&32.49\\
				&&&(+.01)&(+.04)&(-.03)&(+.03)\\
			\end{tabular}
		\end{ruledtabular}
	\end{table}
	
	To address this, we can evaluate the relative contributions to the energy shifts from the change of lattice structure and gap by comparing the results obtained with different strains but identical gaps (by adjusting $U$). In Table~\ref{tab:DFT2}, we show that with 1.0\,\% strain and $U$\,=\,7\,eV, the gap size is comparable to the value obtained with 0.1\,\% strain and $U$\,=\,10\,eV (Table~\ref{tab:DFT1}), but 36\,\% smaller than the value obtained with 1.0\,\% strain and $U$\,=\,10\,eV.
	In agreement with previous discussion, we find that changes in $U$, strain and gap size affect only marginally the B$_{2g}^{(3)}$ and A$_{1g}$-symmetry phonon mode frequencies. On the other hand, the 36\,\% gap size reduction yields a 4\% softening of the B$_{2g}^{(2)}$ mode, whereas the decrease of strain from 1\,\% to 0.1\,\% with constant gap value yields a comparatively larger softening of 10\% frequency of the B$_{2g}^{(2)}$ phonon frequency. 
	
	From these discussions we see that the anomalous behavior of the B$_{2g}^{(2)}$ phonon, both as function of temperature and of strain, is rooted in an incipient instability of the orthorhombic lattice structure against a shear monoclinic distortion of the Ta chains, which is likely to occur irrespective of the electronic gap size. Time-resolved spectroscopy, which has been successfully used in the study of Ta$_2$NiSe$_5$ to disentangle the charge and lattice dynamics~\cite{Fast2017,Fast2018,Fast2021,Fast2023a,Fast2023b}, would provide further support to this view.
	
	Although additional work is required to examine whether this distortion is effectively realized in Ta$_2$NiS$_5$, as suggested by the Raman data, our work demonstrates that the susceptibility towards it is very large, and would likely occur upon compression of the lattice. At the same time, such a perturbation would allow a reduction of the semiconducting gap size, which would offer an interesting way to approach the regime in which an EI can occur (i.e. when the gap size becomes lower than the exciton binding energy).
	
	\begin{table}
		\caption{\label{tab:DFT2}Comparison between the calculated energy values for the B$_{2g}$ phonon modes and band gap under 1.0\,\% strain.}
		\begin{ruledtabular}
			\begin{tabular}{lccc}
				&Gap&B$_{2g}^{(2)}$&B$_{2g}^{(3)}$\\
				\hline
				$U$\,=\,10\,eV&213.5\,meV&17.21\,meV&32.40\,meV\\
				$U$\,=\,7\,eV&136.7\,meV&16.50\,meV&32.48\,meV\\
				Relative change&-36.0\,\%&-4.1\,\%&+0.2\,\%\\
			\end{tabular}
		\end{ruledtabular}
	\end{table}
	
	\section{Conclusion\label{sec:Con}}
	In summary, we have used polarization-resolved Raman spectroscopy to study the response of phononic and electronic degrees of freedom to uniaxial strain in Ta$_2$NiS$_5$. In particular, we found that tensile strain along the $a$ axis results in an anomalously large frequency increase for two B$_{2g}$ phonon modes. The corresponding Gr\"uneisen parameter for the lowest-energy B$_{2g}$ mode is negative and has an magnitude larger than 20. DFT calculations support that such anomalous Gr\"uneisen parameter results from a proximity to a structural instability similar to the one in excitonic insulator candidate Ta$_2$NiSe$_5$. Besides,the band gap of this system increases with strain, as the tensile strain reduces the orbital overlap and in turn the bandwidth. These experimental results establish Ta$_2$NiS$_5$ as a suitable platform to explore strain tuning of lattice dynamics and electronic structure.
	
	\begin{acknowledgments}
		This work was funded by the Deutsche Forschungsgemeinschaft (DFG, German Research Foundation) - TRR 288 - 422213477 (project B03), and Projektnummer 449386310.
		M. F. acknowledges funding from the Alexander von Humboldt fundation, and the YIG preparation program of the Karlsruhe Institute of Technology. R.H. acknowledges support by the state of Baden-W\"{u}rttemberg through bwHPC.
	\end{acknowledgments}
	
	
	%

\end{document}